\documentclass[reprint, amsmath,amssymb,aps,prb,superscriptaddress]{revtex4-1}

\usepackage{graphicx}
\usepackage{dcolumn}
\usepackage{bm}

\begin{document}

\preprint{}

\title{$\mathrm{SU}(N)$ Heisenberg model with multi-column representations}

\author{Tsuyoshi Okubo}
 \email{t-okubo@issp.u-tokyo.ac.jp}
\affiliation{Institute for Solid State Physics, University of Tokyo,
 Kashiwa, Chiba 277-8581, Japan
}%
\author{Kenji Harada}%
\affiliation{Graduate School of Informatics, Kyoto University, Kyoto
 615-8063, Japan}%
\author{Jie Lou}%
\affiliation{Department of Physics, Fudan University, Shanghai 200433, China
}%
\author{Naoki Kawashima}%
\affiliation{Institute for Solid State Physics, University of Tokyo,
 Kashiwa, Chiba 277-8581, Japan
}%
\date{\today}

\begin{abstract}
The $\mathrm{SU}(N)$ symmetric antiferromagnetic Heisenberg model with
 multi-column representations on the two-dimensional square lattice is
 investigated by quantum Monte Carlo simulations. For the representation
 of Young diagram with two columns, we confirm that a valence-bond solid
 order appears as soon as the N\'eel order disappears at $N = 10$
 indicating no intermediate phase. In the case of the representation with
 three columns, there is no evidence for both of the N\'eel and the
 valence-bond solid ordering for $N\ge 15$. This is actually consistent
 with the large-$N$ theory, which predicts that the VBS state immediately
 follows the N\'eel state, because the expected spontaneous order is too
 weak to be detected.
\end{abstract}

\maketitle

Realization of quantum spin liquid in short-range coupling models has
been a popular research target in condensed matter physics for several
decades. One approach to obtain a spin-liquid state is to consider a
Hamiltonian with higher symmetry. Read and Sachdev generalized
antiferromagnetic Heisenberg into $\mathrm{SU}(N)$
symmetry\cite{ReadSachdev1990,ReadSachdev1989}. Based on the $1/N$
expansion they showed that the ground state of the model with
sufficiently large $N$ is a valence-bond-solid (VBS) breaking the
lattice rotational or the translational symmetry spontaneously.
Recently, in terms of the deconfined quantum
criticality\cite{SenthilVBSF2004,SenthilBSVF2004,SenthilBSVF2005}, 
their theory attracts much attention. In particular, the existence of
intermediate state was discussed near the boundary of N\'eel and
VBS\cite{HaradaKT2003,KawashimaTanabe2007,BeachFMS2009}.

\begin{figure}
 \begin{center}
  \includegraphics[scale=0.6]{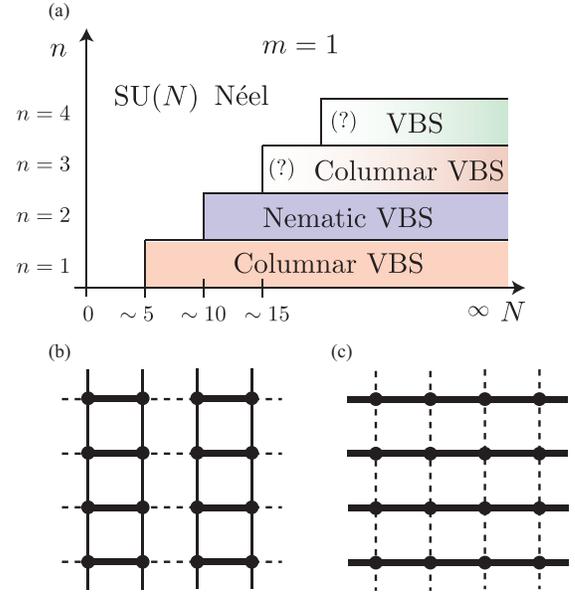} \caption{(color online) (a):
  Schematic phase diagram of the $\mathrm{SU}(N)$ Heisenberg model on
  the square lattice with single-row ($m=1$) representations. The phase
  boundaries for the case of $n=2,3$ are determined in the present
  study, while that for $n=1$ was from
  Ref. \cite{HaradaKT2003,KawashimaTanabe2007}. In the case of $n=3,4$,
  we do not see clear evidence of the spontaneous VBS order in the
  vicinity of the phase boundaries for finite-size QMC simulations. (b),(c): Schematic picture of the columnar VBS (b)
  and the nematic VBS (c) states. Thick solid lines denote larger value
  of $\langle \sum_{\alpha,\beta}^N
  S_i^{\alpha\beta}\tilde{S}_j^{\beta\alpha}\rangle$ while thin solid and dashed
  lines indicate smaller values. }  \label{fig:VBS}
 \end{center} 
\end{figure}

Nature of the ground states of the model can vary depending on the
representation of $\mathrm{SU}(N)$ algebra. Read and Sachdev suggested
that for the model with the representation of $m$ rows and $n$ columns
Young diagram, the ground state phase diagram on the $N$-$n$ plane does
not strongly depends on the value of $m$. Within the $1/N$ expansion,
there are only two types of phases: the small-$N$ N\'eel phase and the
large-$N$ VBS phase (see Fig.~\ref{fig:VBS}(a)). In addition, it was
shown that the nature of the VBS state can be classified by the
remainder of the division of $n$ by 4 on the two-dimensional square
lattice. For $n = 1,3~(\mathrm{mod}~4)$, the VBS state is so called
columnar VBS where both of translational symmetry and $90^\circ$ lattice
rotational symmetry are broken (Fig.~\ref{fig:VBS}(b)). For $n
=2~(\mathrm{mod}~4)$, the VBS state is expected to be a nematic VBS with
breaking only lattice rotational symmetry (Fig.~\ref{fig:VBS}(c)). In the case of $n = 0 ~(\mathrm{mod}~4)$, there
is no spontaneous symmetry breaking, which is an analog of Haldane state
in $S=1$ spin chain.

Beyond the $1/N$-expansion, it was shown that for $(n, m)= (1,1)$ the
ground state is the $\mathrm{SU}(N)$ N\'eel state for $N \le 4$ while it
becomes columnar VBS state for $N \ge 5$ by the quantum Monte Carlo
(QMC) calculation \cite{HaradaKT2003,KawashimaTanabe2007}. Related to
this case, the $\mathrm{SU}(N)$ $J$-$Q$ model was proposed by
Sandvik\cite{Sandvik2007}. The $\mathrm{SU}(N)$ $J$-$Q$ model has an
additional many-body interaction so that the quantum phase transition
between the N\'eel phase and the columnar VBS phase occurs by
continuously changing the Hamiltonian parameter. In order to vary the
Hamiltonian continuously, a continuous-$N$ model was also proposed by
Beach {\it et al.,} \cite{BeachFMS2009}. Because the phase transition
might be a realization of deconfined quantum criticality
\cite{SenthilVBSF2004,SenthilBSVF2004,SenthilBSVF2005}, nature of these
models have attracted much recent interests in condensed matter physics
\cite{Sandvik2007,BeachFMS2009,MelkoKaul2008,KuklovMPST2008,LouSK2009,KaulS2012,
ChenHDKPS2013,HaradaSOMLWTK2013}.

For the case of $n \ge 2$, there were a few studies concerning 
phase boundary between the N\'eel phase and the VBS phase. In QMC
calculation up to $L = 32$ for $L\times L $ square lattice, no evidence
of VBS order was found for $n = 2, 3, 4$ with $m=1$
\cite{KawashimaTanabe2007}. This result appeared to suggest an intermediate
phase between the N\'eel phase and the VBS phases. However, whether the
missing evidence of the VBS order for $n \ge 2$ is due to an intermediate
spin liquid phase or due to the extremely small (but finite) order
parameter beyond numerical limitation has not been clarified up to now.

In this paper, we investigate ground state of the $\mathrm{SU}(N)$
Heisenberg model for $n = 2$ and $3$ with $m = 1$ by using QMC
simulation. The $\mathrm{SU}(N)$ model we considered is an $\mathrm{SU}(N)$ symmetric
antiferromagnetic Heisenberg model on the two-dimensional square lattice
with the periodic boundary condition. Hamiltonian of the model is given
by
\begin{equation}
 \mathcal{H} = \frac{J}{N} \sum_{\langle i,j\rangle,i \in A}
  \sum_{\alpha,\beta=1}^N S_i^{\alpha\beta}\tilde{S}_j^{\beta\alpha},
\end{equation}
where $S_i^{\alpha\beta}, \tilde{S}_j^{\beta\alpha}$ are generators of
$\mathrm{SU}(N)$ algebra, and we consider $J > 0$. On one sublattice $A$
of the lattice, the representation of the generators $S_i^{\alpha\beta}$
is characterized by the Young diagram with a single row ($m=1$) and arbitrary
number $n$ of columns, while we use the conjugate representation
$\tilde{S}_i^{\alpha\beta}$ on the other sublattice. Note that the
conjugate representation satisfies the relation
$\tilde{S}_i^{\alpha\beta} = - S_i^{\beta\alpha}$. We have performed QMC
simulation based on the loop algorithm. We modified ALPS/LOOPER
code\cite{TodoKato2001,ALPS1} for the present purpose\cite{ALPS2}. We
set the inverse temperature $\beta$ as $\beta J= L$ and investigated the
zero temperature properties by extrapolating the results to $L \to \infty$. 

In order to see the VBS orders, we define two order parameters. The
local nematic order parameter is defined as 
\begin{equation}
 \Phi_j \equiv P_{j,y} - P_{j,x},
\end{equation}
where $P_{j,\mu}$ ($\mu = \pm x , \pm y$) is the nearest-neighbor
product of ``magnetic'' moments
\begin{equation}
 P_{j,\mu} \equiv \sum_{\alpha=1}^N S^{\alpha\alpha}_j
  S^{\alpha\alpha}_{j+\bm{e}_\mu}.
\end{equation}
The nematic order parameter characterizes the symmetry breaking of 90
degrees lattice rotation. $\langle \Phi_j\rangle$ takes a finite value for
both of the nematic VBS and the columnar VBS states in the
thermodynamic limit. We also define a local complex order parameter
characterizing the columnar VBS order as
\begin{equation}
 \Psi_j \equiv (-1)^{j_x} \left(P_{j,x} - P_{j,-x}\right) + i(-1)^{j_y} 
  \left(P_{j,y} - P_{j,-y}\right),
\end{equation}
where $j_x$ and $j_y$ are integers representing the lattice coordinates
of site $j$. In the columnar VBS phase $|\langle \Psi_j\rangle| \neq 0$,
while $|\langle \Psi_j\rangle| = 0$ for the N\'eel and the nematic VBS
phases. In order to see the phase transition clearly, we examine the
two-point correlation functions of an observable $O$: $C_O(\bm{R})\equiv
\langle O(\bm{0})O^\dagger(\bm{R})\rangle$. For the N\'eel order, we use the
correlation of a magnetization $S^{\alpha\alpha}_i$:
$C_{\text{Mag}}(\bm{R}) \equiv \sum_{\alpha=1}^{N}
C_{S^{\alpha\alpha}}(\bm{R})$. We also consider the nematic VBS
correlation $C_{\text{Nem}}\equiv C_{\Phi}(\bm{R})$ and the columnar VBS
correlation $C_{\text{Col}}(\bm{R})\equiv C_{\Psi}(\bm{R}).$

\begin{figure}
 \begin{center}
  \includegraphics[scale=0.5]{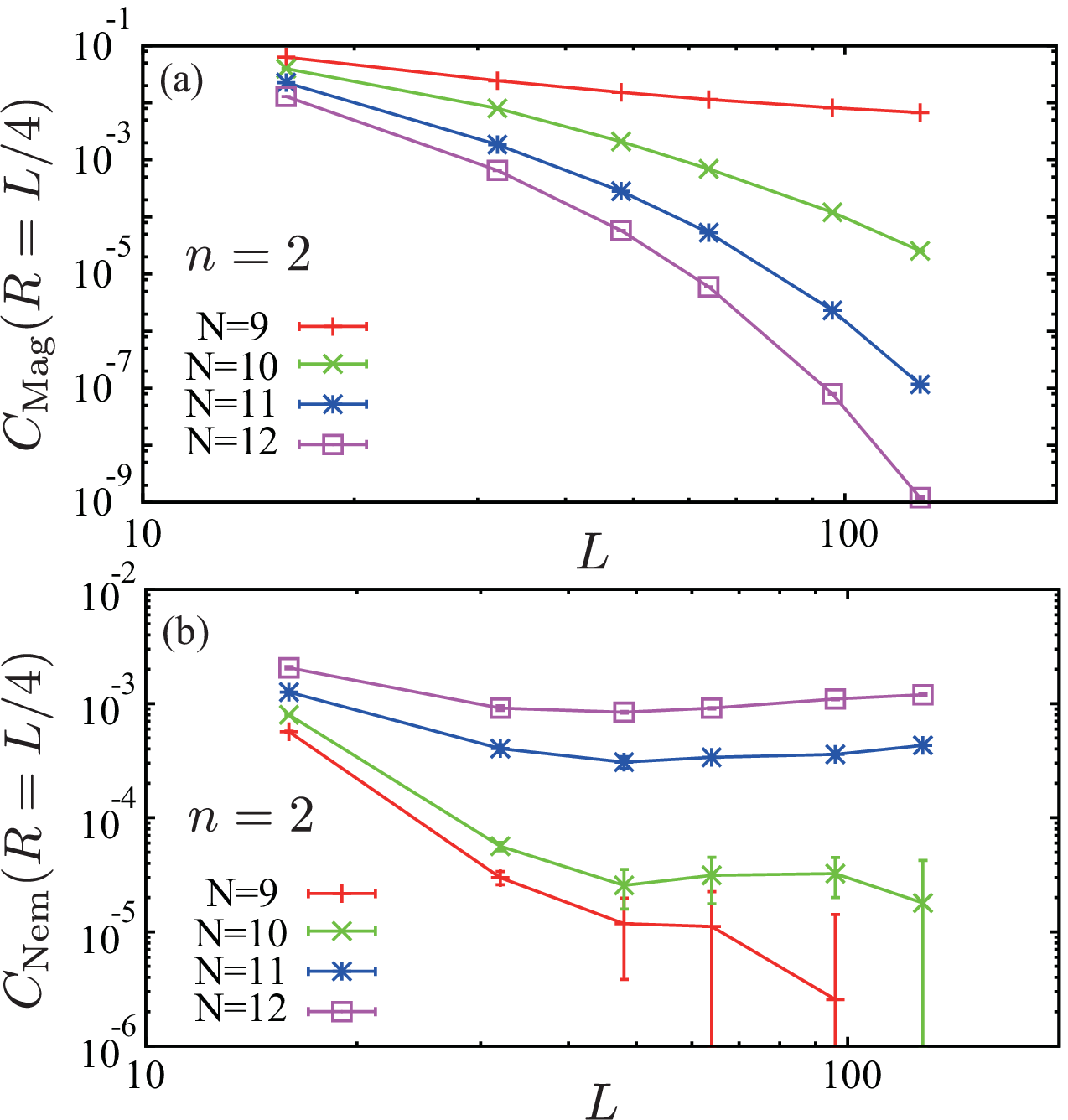}
  \caption{(color online) Log-log plot of the two-point correlation
  functions at $|\bm{R}| = L/4$ for the model with $n=2$ . (a) The
  correlation function of the N\'eel order. (b) The correlation function
  of the nematic order.}  \label{fig:correlation_n2}
 \end{center} 
\end{figure}
\begin{figure}
 \begin{center}
  \includegraphics[scale=0.6]{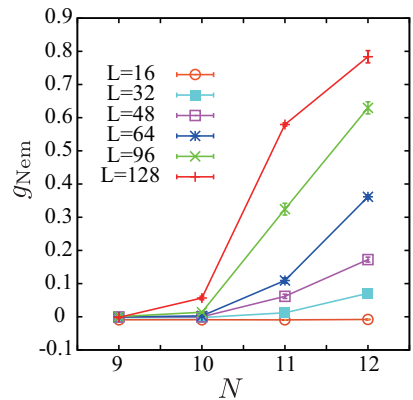}
  \caption{(color online) The binder ratio of the nematic order
  parameter for the model with $n=2$ for $16 \le L \le 128$.}
  \label{fig:binder_n2} 
 \end{center} 
\end{figure}

\begin{figure}
 \begin{center}
  \includegraphics[scale=0.5]{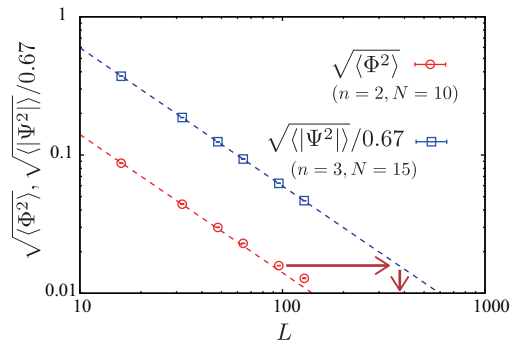} 
  \caption{(color online) Log-log plot of the size dependence of the VBS
  order parameters for $(n,N) = (2, 10)$ and $ (n, N) = (3,15)$. The
  dashed lines represent the fitting curve assuming $1/L$ decay. Arrows
  indicate the correspondence between $n=2$ and $n=3$ cases.}  \label{fig:n2n3_comp}
 \end{center} 
\end{figure}

First, we examine the case of $n=2$. In Fig.~\ref{fig:correlation_n2},
we show the two-point correlations for $n=2$ at $|\bm{R}|=L/4$ for
various sizes $L$ and $N$. For the case of the N\'eel order
(Fig.~\ref{fig:correlation_n2}(a)), we see that the correlation
exponentially decays to zero by increasing $L$ for $N\ge 10$ while it
converges to a nonzero value for $N=9$ indicating that the N\'eel state
is the ground state for $N\le 9$ and it is not for $N\ge 10$. These
observations are consistent with the previous QMC
calculation\cite{KawashimaTanabe2007}. For the nematic order parameter
(Fig.~\ref{fig:correlation_n2}(b)), the two-point correlations tend to
converge to nonzero values for $N \ge 10$, although the situation at
$N=10$ is rather subtle because of larger statistical errors comparable
with the correlation function itself\footnote{At $N=10$ for $L\ge 48$,
estimated standard error is $O(10^{-5})$ which is same order with the
mean value. Thus the converging behavior at $N=10$ is unclear within the
present data set.}.  In order to confirm the appearance of the nematic
VBS order at $N=10$, we plot the Binder ratio of the nematic order
parameter in Fig.~\ref{fig:binder_n2}. The Binder ratio for the nematic
order parameter is given by
\begin{equation}
 g_{\text{Nem}} \equiv \frac{1}{2}\left(3 - \frac{\langle \Phi^4
				   \rangle}{\langle \Phi^2\rangle^2} \right),
\end{equation}
where $\Phi$ is the sum of local nematic order parameters: $\Phi \equiv
L^{-2}\sum_j \Phi_j$. $g_{\text{Nem}}$ is normalized so that
$g_{\text{Nem}} = 0$ for the N\'eel phase, while $g_{\text{Nem}} = 1$
for the nematic (or the columnar) VBS phase. The nematic binder ratio at
$N = 10$ develops as the system size is increased and, as we see in
Fig.~\ref{fig:n2n3_comp}, the order parameter $\sqrt{\langle \Phi^2\rangle}$ slightly deviates upward from the
power-law decay, $\sqrt{\langle \Phi^2\rangle}(L)\propto 1/L$, which should be obeyed
asymptotically when the system is gapped. It indicates the nematic VBS
order at $N=10$ in the thermodynamic limit. We also checked that the
order parameter $\Psi$ shows no evidence of long range order for the case
of $n=2$. From these observations, we conclude that for the case of
$n=2$ the ground state is the N\'eel state for $N \le 9$, while it is
the nematic VBS state for $N \ge 10$. There is no intermediate phase.

\begin{figure}
 \begin{center}
  \includegraphics[scale=0.5]{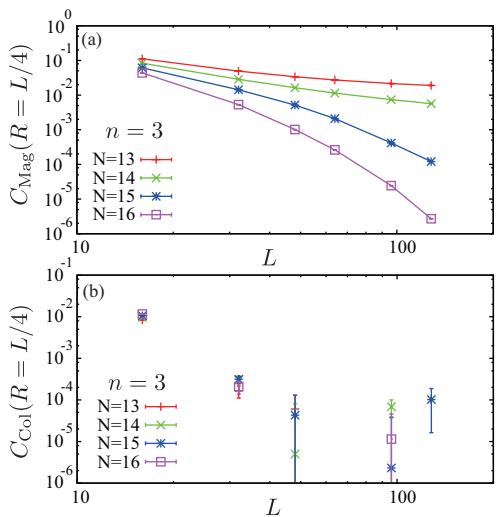}
  \caption{(color online) Log-log plot of the two-point correlation
  functions at $|\bm{R}| = L/4$ for the model with $n=3$ . (a) The
  correlation function of the N\'eel order. (b) The correlation function
  of the columnar VBS order.}  \label{fig:correlation_n3}
 \end{center} 
\end{figure}

Next we move to the case of $n=3$. We plot two-point correlation
functions for $n=3$ at $|\bm{R}|=L/4$ in
Fig.~\ref{fig:correlation_n3}. For the N\'eel order, we clearly see from
the curvature of the curves that the N\'eel state is the ground state
for $N\le 14$, and it is not for $N \ge 15$. On the other hand, we do
not see clear difference among different values of $N$ in the two-point
correlation function of the columnar VBS order (see
Fig.~\ref{fig:correlation_n3}(b)). The behavior of the columnar VBS
order parameters indicates that the VBS order is too small to be visible
even if it exists for $L\le 128$ finite systems with the present
statistical errors. Indeed, based on the $1/N$ expansion Read and
Sachdev proposed that the amplitude of the VBS order parameter becomes
exponentially small by increasing $N$ as $|\langle \Psi_j\rangle| \sim
\exp(-NE_c)$ with the action of a hedgehog
instanton\cite{ReadSachdev1990}. The constant $E_c$ has been calculated as $E_c =
c\ln \xi$ with $c = 0.12459\dots$ in the limit $N\to \infty $ with the spin
correlation length $\xi$ large but fixed\cite{ReadSachdev1989}.

By using the result of the large $N$ theory \cite{ReadSachdev1990},
we try to estimate expected amplitude of the VBS order. For the
columnar VBS order parameter, more precise expression for our
definition of $\Psi_j$ is given by
\begin{equation}
  \left|\langle \Psi_j \rangle\right| = \frac{Na}{\sqrt{2}}\exp(-NE_c),
 \label{eq:Psi_large_N}
\end{equation}
where $a$ is an unknown constant depending on $n/N$. For the nematic VBS
order parameter we also obtain
\begin{equation}
\left|\langle \Phi_j \rangle\right| = \frac{Na}{2}\exp(-NE_c).
 \label{eq:A_large_N}
\end{equation}
We focus on the expected phase boundary of the columnar VBS
phase $N=15$ with $n=3$. The spin correlation length at this parameter
is calculated as $\xi \simeq 5.2$ from a fitting of the correlation
function of the N\'eel order. In the same ratio of $n/N = 0.2$, the spin
correlation length at $N=10$ with $n=2$ is estimated as $\xi \simeq
4.7$. By substituting the values of $\xi$ and $N$ into two equations 
\eqref{eq:Psi_large_N} and \eqref{eq:A_large_N}
with
$E_c \simeq 0.12459 \ln \xi$, we obtain a relation
\begin{equation}
 \left|\langle \Psi_j\rangle_{N=15,n=3}\right| \simeq 0.67 \left|\langle \Phi_j\rangle_{N=10,n=2}\right|.
  \label{eq:Order_parameter_relation}
\end{equation}
In Fig.~\ref{fig:n2n3_comp}, we plot the system size dependence of the
columnar VBS order parameter $\sqrt{\langle |\Psi|^2 \rangle}$, where
$\Psi \equiv L^{-2}\sum_j \Psi_j$, along with that of the nematic VBS
order parameter $\sqrt{\langle \Phi^2 \rangle}$. These order parameters
are expected to converge into $\left|\langle \Psi_j \rangle\right|$ and
$\left|\langle \Phi_j \rangle\right|$, respectively, in the
thermodynamic limit. For the purpose of better comparison, we divide
$\sqrt{\langle |\Psi|^2 \rangle}$ by the factor $0.67$ which appeared in
Eq.~\eqref{eq:Order_parameter_relation}. In the case of $\sqrt{\langle
|\Psi|^2 \rangle}$ it decrease as $L^{-1}$ expected for the case of no
long range order, while the $L$ dependence of $\sqrt{\langle \Phi^2
\rangle}$ changes from $L^{-1}$ around $L\simeq 100$ indicating
development of a weak long range nematic VBS order. From comparison
between $\sqrt{\langle |\Psi|^2 \rangle}$ and $\sqrt{\langle \Phi^2
\rangle}$, we expect that a signature of the columnar VBS order for
$n=3, N=15$ becomes visible for the systems size larger than $L\simeq
400$. Therefore the fact that we did not observe any evidence of the
long range VBS order in the present calculation upto $L=128$ does not
necessarily indicate the presence of intermediate phase where both of
the N\`eel and the VBS order disappears. Because the QMC calculation for
$L \simeq 400$ needs larger computational cost than the available
resources we cannot reach a clear answer for the phase boundary in the
case of $n=3$.

In summary, we have investigated the ground state property of an
$\mathrm{SU}(N)$ symmetric antiferromagnetic Heisenberg model on the
two-dimensional square lattice for the representations with the $n =2$
and $n=3$ columns Young diagrams. For $n=2$, we conclude that the N\'eel
state is the ground state for $N\le 9$ while the nematic VBS state
becomes the ground state for $N\ge 10$. Thus, there is no intermediate
state between them (see Fig.~\ref{fig:VBS}(a)). For $n=3$, the ground state
for $N\le 14$ is the N\'eel state and it disappears for $N\ge
15$. Although we observed no evidence of the expected columnar VBS order
for $N\gtrsim 15$, this observation does not exclude the columnar VBS
order in this case, because we estimated that the signature of the VBS
order was invisible for smaller sizes $L\lesssim 400$ even if it
eventually converges to a finite value. Determining the VBS phase
boundary for $n\ge 3$ requires further studies. Our analysis indicates
that we need careful extrapolations of the finite-size data into the
thermodynamic limit. Naive extrapolations may lead to an incorrect
characterization of the intermediate region
even if a weak VBS order is eventually stabilized\cite{Sandvik2012}.

\begin{acknowledgements}
 We owe helpful discussions to S.~Todo, T.~Suzuki, H.~Watanabe, and
 H.~Matsuo. 
 The computation in the present work is partly executed on computers as
 the Supercomputer Center, ISSP, University of Tokyo. The present work
 is financially supported by MEXT Grant-in-Aid for Scientific Research
 (B)(25287097) and (C)(26400392), and by CMSI, MEXT-SPIRE, Japan.

\end{acknowledgements}

\bibliography{SUN_multi}

\end{document}